\documentclass{sig-alt-wbl} 
\usepackage{epsfig}
\usepackage{latexsym}
\usepackage{hyperref} 

\catcode`\.=\active\def.{\char'56\allowbreak}\catcode`\.=12
\catcode`\/=\active\def/{\char'57\allowbreak}\catcode`\/=12
\catcode`\:=\active\def:{\char'72\allowbreak}\catcode`\:=12  
\catcode`\@=\active\def@{\char'100\allowbreak}\catcode`\@=12 
\catcode`\-=\active\def-{\char'55\allowbreak}\catcode`\-=12  
\def\URL{\bgroup\catcode`\.=\active\catcode`\/=\active\catcode`\:=\active\catcode`\@=\active\catcode`\-=\active\def~{\char126}\tt\URLaux}
\def\URLaux#1{#1\egroup}

\newlength{\halfwidth}
\begin{document}

\title{
RNAnet a Map of Human Gene Expression\\
\href{http://bioinformatics.essex.ac.uk/users/wlangdon/rnanet/}
{\tt {http://bioinformatics.essex.ac.uk/users/wlangdon/rnanet/}}
\mbox{ }
\\[1ex]
\normalsize
This web site is free and open to all users and there is no login requirement
\vspace*{-4ex}
}
\numberofauthors{1} 
\author{
{{W. B. Langdon}\,$^{1,*}$},
Olivia {Sanchez Graillet}\,$^{2}$ and A. P. Harrison\,$^2$%
\\[1em]
\affaddr{
$^{1}$
Computer Science, 
        King's College, London,
UK
}
\email{William.Langdon@kcl.ac.uk}
\\
\affaddr{
$^{2}$
Departments of
Mathematical and Biological Sciences,
University of Essex
}
\email{\{osanch,harry\}@essex.ac.uk}
\\[1ex]
{\normalsize
Keywords: Systems Biology, exon coexpression, Bio-computing,
integrating disparate data sources
}
}

\bibliographystyle{unsrt}

\maketitle

\noindent
\href{http://bioinformatics.essex.ac.uk/users/wlangdon/rnanet/}
{RNAnet} provides a bridge between two widely used
Human gene databases.
Ensembl 
describes DNA sequences and transcripts
but not experimental gene expression.
Whilst GEO 
contains actual expression levels from many thousands of Human samples
but,
although samples in GEO can be queried by experiment,
comparison across the whole of GEO is not supported.
RNAnet provides immediate access to 
thousands of Affymetrix HG-U133~2+ 
measurements provided by GEO
covering 
Human genes in most medically interesting tissues.
Data have been quantile normalised and scanned for a
variety of common GeneChip errors.
There are copious links back into
GEO and Ensembl.

Without RNAnet comparison across experiments in GEO is very labour
intensive requiring down loading and manual cleaning of 
data files for each microarray in each experiment.
Previously normalising more than a few dozens GeneChips was tricky.
Having downloaded tens of thousands of microarray datasets,
before RNAnet, we could normalise all HG-U133~2+ in ten hours.
With RNAnet anyone can access cleaned quantile normalised data in seconds.

Further, since we have data from across many different tissues and
medical conditions GEO data 
can be used
to find patterns of co-expression.
RNAnet shows that the network of
strong correlations is huge but sparse.
Thousands of genes interact strongly with thousands
of others.
Conversely, 
tens of thousands of genes 
interact
strongly with less than 100 others.
I.e.\
RNAnet
gives new views for RNA Systems Biology.
It builds on free but very valuable databases.

\section{Using Normalised GEO Data}
\label{sec:crosshairs}
\noindent
The URL 
\href{http://bioinformatics.essex.ac.uk/users/wlangdon/rnanet/probeset.php?1556291_at}
{\tt{http://bioinformatics.essex.ac.uk/users/}}
\linebreak
\href{http://bioinformatics.essex.ac.uk/users/wlangdon/rnanet/probeset.php?1556291_at}
{\tt{wlangdon/rnanet/probeset.php?}}
followed by an Affymetrix probeset identifier
(e.g.\ 1556291\_at)
immediately gives a table of all the log$_e$ normalised values
for the
probeset.
Exceptional or suspect values are flagged with a ``?''.
The table can be loaded into a spread sheet or other analysis tools.
``Perfect match'' data (PM) 
have a clickable label 
pointing to their GEO  experiment.
``Mis-match'' data (MM) have similar hyperlinks which take you
directly to the GEO description of the individual tissue sample.
 
Users of Firefox can interactively plot data,
either from the same or different probesets
using
\href{http://bioinformatics.essex.ac.uk/users/wlangdon/rnanet/scatter.html#1570561_at.pm1,1570561_at.pm3}
{\tt{http://bioinformatics.}}
\href{http://bioinformatics.essex.ac.uk/users/wlangdon/rnanet/scatter.html#1570561_at.pm1,1570561_at.pm3}
{\tt{essex.ac.uk/users/wlangdon/rnanet/scatter.html}}.
The crosshairs provides access to individual values
and hyperlinks into the metadata held by GEO\@.
Probes can either be specified by Affymetrix probeset id or
Ensembl 
\href{http://bioinformatics.essex.ac.uk/users/wlangdon/rnanet/scatter.html#ENSE00000891612.pm,ENSE00000664794.pm}
{exon id}
or a mixture.
Colour allows multiple plots on the same graph.

\section{Interactive Correlation Heat maps}
\label{sec:heapmap}
\noindent
The 
\href{http://bioinformatics.essex.ac.uk/users/osanch/heatmaps/version6/}
{web site}
contains many tens of thousands of pre-calculated heat maps
for Mouse,
Arabidopsis,
Rice and Soybean, as well as Human.
For Human genes, RNAnet also supports flexible interactive
construction of correlation 
\href{http://bioinformatics.essex.ac.uk/users/wlangdon/rnanet/correlation.html#ENSE00000891612,ENSE00000664794}
{heat maps}.
Any set of probes can be correlated either if the probes map in sense,
antisense or in both directions to the exon. The probes can be from
the same or different probesets and the heat maps can be of any size.
Typically $10 \times 10$ correlations takes
about a second to calculate and display.
Again GeneChip data may be requested either by Affymetrix
\href{https://www.affymetrix.com/analysis/netaffx/}
{probeset} 
or 
Ensembl 
\href{http://bioinformatics.essex.ac.uk/users/wlangdon/rnanet/gene_description.html}
{exon} id.
(Usually several Affymetrix probes measure an exon.
The one chosen as being typical, i.e.\ most correlated,
is indicated with an asterisk~*.)
Correlations follow the same 
\href{http://bioinformatics.essex.ac.uk/users/wlangdon/colour.html}
{colour coding}
as the fixed matrices.
Hyperlinks on the matrix lead to the underlying scatter plot.
(Remember to press 
\href{http://bioinformatics.essex.ac.uk/users/wlangdon/rnanet/scatter.html#209396_s_at.pm9,209396_s_at.pm1}{plot}.)
Additionally the text button displays the 
averages and correlations in numeric form.

\section{RNA Systems Biology}

\noindent
We calculated the correlations across all of GEO of 24\,132 exons
with each other.
The main RNAnet graphical 
\href{http://bioinformatics.essex.ac.uk/users/wlangdon/rnanet/rnanet.html}
{screen}
allows Firefox users to query
these 290 million correlation coefficients
by gene name or Ensembl exon id.
Strong correlations or anti-correlations can be plotted on
a PCA analysis of the 290 million correlations.
Additionally up to ten exons closest to the dragable crosshairs
(cf.\ sect.~\ref{sec:crosshairs})
can be displayed.
Once an exon is selected it should be locked into the display
(2$^{nd}$ box on right) to avoid the next search over writing it. 
Again heat maps are created and displayed as needed.
Due to non-unique mappings between Ensembl exons and Affymetrix
probesets and concerns about sequence quality,
correlations for only 24\,132 Ensembl exons are available.
They represent about half the Human genes.

To Dec 2009, 
\href{http://my6.statcounter.com/project/standard/stats.php?account_id=1905560&login_id=5&code=aabd35e5d705984afb4bed4ca2280b15&guest_login=1&project_id=3324720}
{3\,585} pages had been loaded by people 
(excluding Essex and King's).
In the last 15 months there have been $\approx 250$
down loads per month.
RNAnet was publicised at UK Affy 2008
and a poster presented at 
\href{http://www.cs.ucl.ac.uk/staff/W.Langdon/WBL_papers.html#langdon:2008:EMBL}
{EMBL~2008}. 
Cf.\ technical report
\cite{CES-486}.
RNAnet was used to corroborate experimental results
on Mycoplasma contamination
\cite{Astarloa:2009:BT}.

\vfill

\bibliography{references,gp-bibliography}

\begin{thebibliography}{1} 

\bibitem{CES-486}
W.~B. Langdon.
\newblock A map of human gene expression.
\newblock 
\href{http://www.essex.ac.uk/dces/research/publications/technicalreports/2008/CES-486.pdf}
{CES-486},
 University of Essex,
 UK, July 2008.

\bibitem{Astarloa:2009:BT}
\href{http://www.biotechniques.com/BiotechniquesJournal/2009/December/Letter-to-the-editor-Unexpected-presence-of-mycoplasma-probes-on-human-microarrays/biotechniques-181035.html}
{E.~Aldecoa-Otalora}, {\em et al.}
\newblock Unexpected presence of Mycoplasma probes on human microarrays.
\newblock {\em BioTechniques}, 47(6), pp1013--1016, December 2009.

\end{thebibliography}

\end{document}